\documentclass[twocolumn,amsmath,amssymb]{revtex4}
\usepackage{bm}
\usepackage{graphicx}
\usepackage{epstopdf}

\begin{document}

\newcommand{\pp}[1]{\phantom{#1}}
\newcommand{\be}{\begin{eqnarray}}
\newcommand{\ee}{\end{eqnarray}}
\newcommand{\Sinn}{\textrm{Sin }}
\newcommand{\Coss}{\textrm{Cos }}
\newcommand{\Sin}{\textrm{Sin}}
\newcommand{\Cos}{\textrm{Cos}}

\title{
Arithmetic loophole in Bell's theorem: Overlooked threat to entangled-state quantum cryptography
}
\author{Marek Czachor}
\affiliation{
Zak{\l}ad Fizyki Teoretycznej i Informatyki Kwantowej,
Politechnika Gda\'nska, 80-233 Gda\'nsk, Poland
}

\begin{abstract}
Bell's theorem is supposed to exclude all local hidden-variable models of quantum correlations. 
However, an explicit counterexample shows that a new class of local realistic models, based on generalized arithmetic and calculus, can exactly reconstruct rotationally symmetric quantum probabilities typical of two-electron singlet states. Observable probabilities are consistent with the usual arithmetic employed by macroscopic observers, but counterfactual aspects of Bell's theorem are sensitive to the choice of hidden-variable arithmetic and calculus. The model is classical in the sense of Einstein, Podolsky, Rosen, and Bell: elements of reality exist and probabilities are modeled by integrals of hidden-variable probability densities. Probability densities have a Clauser-Horne product form typical of local realistic theories. However, neither the product nor the integral nor the representation of rotations  are  the usual ones. 
The integral has all the standard properties but only with respect to the  arithmetic that defines the product. Certain formal  transformations of integral expressions one finds in the usual proofs \`a la Bell do not work, so standard Bell-type inequalities cannot be proved.   The system we consider is deterministic, local-realistic, rotationally invariant, observers have free will, detectors are perfect,  hence the system is free of all the canonical loopholes discussed in the literature. 
\end{abstract}
\maketitle

\section{Introduction}

The problem posed by Einstein, Podolsky, and Rosen \cite{EPR,Bohm0}, and reformulated by Bell \cite{Bell}, is as follows: Can there exist elements of reality whose knowledge would allow to predict  in advance results of quantum measurements? The advent of quantum cryptography  \cite{BB,Ekert,BBM92,Gisin} has turned the purely academic debate into a practical one: any loophole in Bell's reasoning creates a potential threat for security of data transmission.

Bell's inequality \cite{Bell} does not apply to systems that do not satisfy at least one assumption needed for its proof. 
This includes nonlocal hidden variables \cite{Bohm}, theories based  on detector inefficiency \cite{Pearle}, locally incompatible random variables \cite{Fine,MC88,Wolf,Kup,Kup2}, observers with limited freedom of choice \cite{t'Hoft}, and contextual cognitive models \cite{Kh,Aerts}. In each of these cases it is easy to understand why the inequality cannot be derived. 
Detector inefficiency was used to hack a Bell-type cryptosystem a long time ago \cite{hack0,hack}. Threats based on nonlocal hidden variables, as well as remedies against them,  are less known \cite{ACP}.

At the other extreme one finds various abstract constructions, involving probability manifolds \cite{Gudder}, non-measurable sets \cite{Pitowski}, or non-computable fractals \cite{Palmer}.  However, the more abstract the model, the more controversial and obscure its physical and probabilistic interpretation. 

What I will discuss is much more down to earth.
Quite recently I have identified a new, `arithmetic'  loophole in the proof of the theorem \cite{Czachor2020}. It remained to construct an explicit  counterexample, simultaneously free of all the other loopholes discussed in the literature. The article shows how to do it. The observers have free will, detectors are ideal, hidden variables are local, and yet the derived probabilities are exactly those implied by quantum mechanics. 

The trick is in the unexplored mathematical freedom: the form of hidden-variable arithmetic and calculus. Arithmetic is a natural language of mathematics. 
It defines the ways we add, subtract, multiply, and divide numbers.  Modified arithmetic implies a modified calculus. However, as there are different languages, there exist different arithmetics and calculi.  The same set of physical variables may be equipped with several coexisting arithmetics. In the context of Bell's theorem this will lead to formulas of the form
\be
N_{kl}/N
=
\int\chi_{\alpha k}^1(\lambda)\odot \chi_{\beta l}^2(\lambda)\odot
\rho(\lambda){\rm D}\lambda.\label{intro}
\ee
Here the left-hand side is the number obtained in experiment. The right-hand side is the theoretical prediction. 
Symbols such as $/$ (division employed by experimentalists to process actual data) or $\odot$ (multiplication employed by theorists to perform counterfactual calculations) correspond to different but mutually consistent arithmetics. The fact that formulas such as (\ref{intro}) are mathematically possible is a manifestation of the consistency. Bell's inequality becomes a macroscopic test for microscopic arithmetic. 

Locality, the key assumption of Bell, effectively means that probabilities (\ref{intro}) have a {\it product\/} form, a  notion that depends on arithmetic. So, how many arithmetics are available if we assume that probabilities are represented by non-negative real numbers summing to 1? The answer may be surprising: infinitely many! It remains to find a correct hidden-variable arithmetic (and calculus it implies) and prove that it predicts local-realistic probabilities that are identical with the quantum ones. The resulting hidden-variable  model of probability is perhaps not exactly classical, but is based on Einstein's elements of reality so is sufficiently classical to create a problem for quantum cryptography, even in its most ideal device-independent version \cite{Acin}. 

Since the subject is unknown to a wider audience, we will gradually develop the construction.
We will begin with arithmetic of parallel-connected resistors. Although the system is well understood from a physical point of view, its arithmetic aspects may appear paradoxical. In particular, there is  a nontrivial relation between addition and multiplication, a fact with consequences for natural numbers.

The next example is related to the problem of dark energy. We will see that accelerated expansion of the Universe can be regarded as a consequence of a mismatch between two arithmetics: the one we normally use, and the one applying to cosmological-scale observers \cite{MCdark2}. The example is particularly relevant for our discussion. It shows that `large' and `small' systems may be in principle based on different types of arithmetic. In  the context of Bell's theorem it is us, the macroscopic-scale observers who are `large', while the hidden-variables are `small'.

Finally, we construct the hidden-variable model of singlet-state correlations. Technically it is  based on two elements: The product which defines locality in Clauser-Horne-type probabilities \cite{CH,CH2}, and the integral which relates hidden variables with observable averages. 
Our model is further analyzed from a geometric perspective. We will see that it is rotationally symmetric, a property one expects from singlet state correlations, but this rotational symmetry is as hidden as the hidden variables themselves. 

The construction is simple, one just has to get used to a more general perspective, whose unifying and generalizing power only starts to be appreciated by scientific community \cite{jizba,entropy}. I believe the proposed formulation circumvents all the basic limitations imposed by Bell's theorem. Most importantly, the model is probabilistically quantum enough to fake quantum correlations, and classical enough to allow for eavesdropping in quantum cryptography.

\section{The world according to  Resistor}
\label{Sec2}

Let us begin with the example that is truly down to earth and easy to understand. 
A parallel configuration of resistors is a resistor whose resistance is computed by means of the harmonic addition
\be
R_1\oplus R_2
=
\frac{1}{1/R_1+1/R_2}
=
f^{-1}\big(f(R_1)+f(R_2)\big),\label{R1+R2}
\ee
where $f(x)=f^{-1}(x)=1/x$.  An analogously defined multiplication remains unchanged, 
\be
R_1\odot R_2
&=&
f^{-1}\big(f(R_1) f(R_2)\big)\\
&=&
\frac{1}{(1/R_1)(1/R_2)}=R_1R_2.
\ee
If we add $n$-times the same resistance $R$ we obtain
\be
\underbrace{R\oplus\dots\oplus R}_n =R/n.\label{R/n}
\ee
Although the physical meaning of  (\ref{R/n})  is obvious, it suggests that $\oplus$ is not an addition in the ordinary sense of the word. Indeed,
\be
\underbrace{R\oplus\dots\oplus R}_n \neq n\odot R.\label{n.R}
\ee
Apparently, the new arithmetic operations,  $\oplus$ and $\odot$, are mutually inconsistent. On the other hand, however, it is clear that 
\be
f(R_1\oplus R_2) &=& f(R_1)+ f(R_2),\\
f(R_1\odot R_2) &=& f(R_1)f(R_2),
\ee
and thus $f$ makes  the `parallel arithmetic' isomorphic to the standard arithmetic of $\mathbb{R}$ (we only have to be cautious at 0). $\oplus$ and $\odot$ are commutative and associative, and $\odot$ is distributive with respect to $\oplus$. So how is that two mathematically isomorphic structures cannot play the same mathematical roles?

In fact, they {\it can\/} play the same roles. The problem is with the meaning of $n$. The natural number $n$ at the right-hand side of (\ref{n.R}) is {\it not\/} a natural number in the sense of the new arithmetic. In order to understand why,  we first  have to clarify what should be meant by `zero' and `one'. Once we define a `one' we can add it several times to itself. The result should be a well defined natural number. 

`Zero' is an element $0'$ such that $x\oplus 0'=x$ for any $x$. An insulated wire is a parallel configuration of resistors with  insulation in the role of an infinitely resistant resistor. Insulation does not influence the wire, $R\oplus\infty=R$,  hence $\infty=0'$. `One' is an element $1'$ such that $x\odot 1'=x$, but since multiplication is unchanged we get  $1'=1$. Greater natural numbers are constructed iteratively,
\be
2'&=&1'\oplus 1'=f^{-1}\big(f(1')+f(1')\big)=f^{-1}(2)=1/2,\\
3'&=&2'\oplus 1'=1'\oplus 1'\oplus 1'=f^{-1}(3)=1/3,\\
&\vdots&\nonumber\\
n'&=&=(n-1)'\oplus 1'=1'\oplus \dots\oplus 1'=f^{-1}(n)=1/n.\nonumber\\
\ee
Accordingly, $n'=f^{-1}(n)=1/n$ is the harmonic representation of $n$. More precisely, $n'$ {\it is\/} the natural number from the point of view of the harmonic arithmetic. Alternatively, following Benioff \cite{B2002,B2005,B2005b}, we could say that $f^{-1}$ is a  {\it value function\/} which maps a natural number $n$ into its value. Benioff's natural numbers are just abstract elements of a well ordered set and in themselves do not possess concrete values. The latter are produced by value functions. The natural number $n'$ satisfies the consistency condition
\be
n'\oplus m' &=& (n+m)'
\ee
as one can directly verify by inserting $n'=1/n$ and $m'=1/m$ into (\ref{R1+R2}). The same rules apply to $1'=f^{-1}(1)$ and $0'=f^{-1}(0)=\lim_{x\to 0_+}f^{-1}(x)$. As we can see, the harmonic multiplication actually  {\it is\/} a repeated addition:
\be
\underbrace{R\oplus\dots\oplus R}_n = n'\odot R=n'R.\label{n.R'}
\ee
Subtraction and division are defined analogously,
\be
R_1\ominus R_2
&=&
f^{-1}\big(f(R_1)- f(R_2)\big)\\
&=&
\frac{1}{(1/R_1)-(1/R_2)},\\
\ominus R
&=&
0'\ominus R
=\frac{1}{0-(1/R)}=-R,
\ee
with the convention that $R\ominus R=\infty=0'$;
\be
R_1\oslash R_2
&=&
f^{-1}\big(f(R_1)/ f(R_2)\big)\\
&=&
\frac{1}{(1/R_1)/(1/R_2)}=R_1/R_2.
\ee
The new arithmetic involves an ordering relation: 
$x\le'y$ if and only if $f(x)\le f(y)$. In particular, $r'\le' s'$ if and only if $r\le s$. 
The 6-tuple $\{\mathbb{R},\oplus,\ominus,\odot,\oslash,\le'\}$ defines an arithmetic which, in the terminology of Burgin \cite{Burgin77,Burgin2010,BC}, is a non-Diophantine projective arithmetic with projection $f$ and coprojection $f^{-1}$. 

A frequentist definition of probability parallels the standard one (the number $n'$ of successes divided by the number $N'$ of trials),
\be
p' &=& n'\oslash N'=n'/N'=f^{-1}(n)/f^{-1}(N)
\nonumber\\
&=&
N/n=f^{-1}(n/N).
\ee
Probabilities sum to one because 
\be
n'\oslash N'\oplus (N'\ominus n')\oslash N'=1'=1.
\ee
Despite  appearances, $p'=N/n$ is {\it not\/} greater than one --- not in the new arithmetic. Indeed,
$p'>' 1'=1$ if and only if $n/N=f(p')>  f(1')=1$, which is impossible. 

Anyone for whom this paper is a first encounter with non-Diophantine arithmetic should pause here and contemplate the result. The notions of `greater' or `smaller'  are local concepts.  Just like `above' and `below' in the antipodic cities of Auckland and Seville. There are many analogies between non-Diophantine arithmetics and non-Euclidean geometries. Something which is larger in one arithmetic may appear smaller in another one (e.g. $0'=\infty$). A number which is negative in one arithmetic can be positive in another one (the arithmetic in $\mathbb{R}_+$, defined by $f(x)=\ln x$, implies $\ominus x=1/x\in \mathbb{R}_+$). However, in order not to confuse the reader, it should be stressed that in the hidden variable model we will discuss below the two arithmetics will  involve the same ordering relation: $\le'$ will be equivalent to $\le$. The loophole will technically follow from Diophantine non-linearity (i.e. non-Diophantine linearity) of hidden-variable integrals.

The two arithmetics are exactly symmetric with respect to each other: $x'=f^{-1}(x)=1/x$ implies $x=f^{-1}(x')=1/x'$, 
\be
x\oplus y &=& f^{-1}\big(f(x)+f(y)\big)
\ee
implies 
\be
x+y &=& f^{-1}\big(f(x)\oplus f(y)\big).
\ee
Which of the natural numbers, $n'=1/n$ or $n=1/n'$, are those we learned as kids? Everything in one arithmetic is exactly upside-down in the other one.  Maybe it is {\it us\/} who live in the Matrix of wires and resistors?
There is absolutely no criterion telling us which of the two arithmetics is Diophantine. This relativity of arithmetics will become essential for the reformulation of the problem of dark energy we will give in the next section.

Non-Diophantine arithmetics imply non-Newtonian calculi \cite{BC,GK,G79,G83,Pap1993,Pap2008,G,MC2015,ACK2016a,ACK2016b,ACK2018,Czachor2019}, in this concrete example a harmonic one.
A harmonic derivative of a function $A:\mathbb{R}\to \mathbb{R}$ is defined in the usual way but by means of the harmonic arithmetic,
\be
\frac{{\rm D}A(x)}{{\rm D}x}
&=&
\lim_{\delta\to 0'}
\big(A(x\oplus\delta)\ominus A(x)\big)\oslash \delta\label{DA00}\\
&=&
\lim_{\delta\to \infty}
\frac{A(x\oplus\delta)\ominus A(x)}{\delta}.
\ee
The derivative is a linear map and satisfies the Leibniz rule (both properties defined with respect to $\oplus$ and $\odot$). It was introduced and studied in \cite{GK}, and then rediscovered at least twice \cite{Pap1993,MC2015}. 
Non-Newtonian calculus leads to unique non-Diophantine  generalizations of all the functions defined by means of derivatives or integrals.  For example, one can directly check that $A(x)=e^{-1/x}$ is the harmonic exponential function, i.e. satisfies
\be
\frac{{\rm D}A(x)}{{\rm D}x}=A(x),\quad A(0')=1',
\ee
and $A(x\oplus y)=A(x)\odot A(y)$. 
Rewriting $e^{-1/x}$ as
\be
A(x)=f^{-1}\big(e^{f(x)}\big)
\ee
we can understand why $A(x)$ plays a role analogous to Newtonian $e^x$. Continuing in a similar vain, we will  arrive at a full calculus, linear algebra, or probability theory. Actually, all of physical theories will have their harmonic analogues. 

Before we will formulate a non-Diophantine/non-Newtonian version of sub-quantum hidden variables, let us first have a look at another, in a sense dual problem of cosmological-scale arithmetic \cite{MCdark2}. The example clearly shows that `small' and `large' systems may work with different arithmetics which are nevertheless defined in the same set of real numbers.

\section{Small-observer perspective: Dark energy as a problem of arithmetic}

Friedman equation for a dimensionless scale factor evolving in a dimensionless time \cite{Hartle}, 
\be
\frac{{\rm d}a(t)}{{\rm d}t}
=
\sqrt{\Omega_\Lambda a(t)^2 +\frac{\Omega_M}{ a(t)}},\quad a(t)>0,\label{FR1}
\ee
is exactly solvable,
\be
a(t)
=
\left(
\sqrt{\frac{\Omega_M}{\Omega_\Lambda}}\sinh \frac{3\sqrt{\Omega_\Lambda}(t-t_1)}{2}
\right)^{2/3},\quad t> t_1.\label{a(t)1}
\ee
The dimensionless time is here expressed in units of the Hubble time $t_H\approx 13.58\times 10^9$~yr. 
It correctly models the observed cosmological expansion if  $\Omega_M=0.3$, $\Omega_\Lambda=0.7$ \cite{DE1,DE2}.
Following  \cite{MCdark2} we will now show that (\ref{a(t)1}) can be obtained with $\Omega_\Lambda=0$, if we change the arithmetic of time. The logic of the example is similar to the argument that potential forces can be replaced by free geodesic motions in curved geometries. 

The standard Diophantine/Newtonian Friedman equation without $\Omega_\Lambda$,
\be
\frac{{\rm d}a(t)}{{\rm d}t}
=
\sqrt{\frac{\Omega_M}{ a(t)}},\quad a(t)>0,\label{FR11}
\ee
is first rewritten in a general non-Diophantine/non-Newtonian form,  not specifying the arithmetics of $\mathbb{X}\ni t$ and $\mathbb{Y}\ni a(t)$, namely 
\be
\frac{{\rm D}a(t)}{{\rm D}t}
=
\Omega_M^{(1/2)_\mathbb{Y}} \oslash_\mathbb{Y} a(t)^{(1/2)_\mathbb{Y}},\quad
a(t)>_\mathbb{Y}0_\mathbb{Y},\label{F}
\ee
where $a^{(1/2)_\mathbb{Y}}\otimes_\mathbb{Y} a^{(1/2)_\mathbb{Y}}=a$, i.e.
\be
a^{(1/2)_\mathbb{Y}}=f^{-1}_\mathbb{Y}\left(\sqrt{f_\mathbb{Y}(a)}\right).
\ee
All the arithmetic operations in $\mathbb{X}$ and $\mathbb{Y}$ are induced from the usual (Diophantine) arithmetic of  $\mathbb{R}$ by means of one-to-one maps $f_\mathbb{X}:\mathbb{X}\to \mathbb{R}$, $f_\mathbb{Y}:\mathbb{Y}\to \mathbb{R}$, in exact analogy to the harmonic arithmetic discussed in the previous section. 
(\ref{F}) is solved by \cite{MCdark2}
\be
a(t)
&=&
f_\mathbb{Y}^{-1}
\Big(
\big(3f_\mathbb{Y}\big(\Omega_M^{(1/2)_\mathbb{Y}}\big)f_\mathbb{X}(t)/2\big)^{2/3}
\Big).\label{A(t)}
\ee
Its comparison with (\ref{a(t)1}), written as
\be
a(t)
&=&
\left[
\frac{3}{2}\sqrt{\Omega_M}\frac{2}{3\sqrt{\Omega_\Lambda}}\sinh \frac{3\sqrt{\Omega_\Lambda}}{2}(t-t_1)
\right]^{2/3},
\ee
suggests a linear $f_\mathbb{Y}(y)=\lambda y$. Inserting $f_\mathbb{Y}\big(\Omega^{(1/2)_\mathbb{Y}}\big)=\sqrt{f_\mathbb{Y}(\Omega)}=\sqrt{\lambda\Omega}$ into (\ref{A(t)}),
\be
a(t)
&=&
\lambda^{-1}
\big(3\sqrt{\lambda\Omega_M}f_\mathbb{X}(t)/2\big)^{2/3}
\\
&=&
\big(3\lambda^{-1}\sqrt{\Omega_M}f_\mathbb{X}(t)/2\big)^{2/3},
\ee
we arrive at 
\be
f_\mathbb{X}(t)
&=&
\frac{2}{3\sqrt{\Omega_\Lambda}}\sinh \frac{3\sqrt{\Omega_\Lambda}}{2}(t-t_1)\label{43}\\
&\approx& 
0.8\, \sinh \frac{t-t_1}{0.8},\label{0.8...}\\
f^{-1}_\mathbb{X}(r)
&=&
t_1+\frac{2}{3\sqrt{\Omega_\Lambda}}\sinh^{-1} \frac{3\sqrt{\Omega_\Lambda}}{2}r,\\
0_\mathbb{X} &=& f^{-1}_\mathbb{X}(0)=t_1,\\
\lambda
&=& \sqrt{\Omega_M/0.3}.
\ee
Assuming $\Omega_M=1$ we find $\lambda=1.82574$. $\lambda\neq 1$ can be incorporated into a change of units as $a(t)$ is here dimensionless.

Cosmological-scale observers, who employ their own arithmetic related by (\ref{0.8...}) to the arithmetic we are taught at school, believe the Universe at {\it their\/} scales expands according to Einstein's general relativity with zero cosmological constant. But they are aware of the dark energy problem: Small objects, such as galaxies or planetary systems, expand with unexplained deceleration...

\section{Large-observer perspective: Bell's theorem as a problem of arithmetic}

Note that in the dark energy example the macroscopic observers are small if compared with the observed system (the Universe). In the hidden-variable problem the macroscopic observers are large. This  instructive `duality'  helps us to switch between two perspectives. In principle, we can imagine that hypothetical sub-quantum observers are aware of a Bell-type theorem stating that macroscopic observers (that is --- us) cannot exit as elements of reality.

Let us now formulate a local hidden-variable theory of the Einstein-Podolsky-Rosen-Bohm two-electron singlet-state correlations.  
The resulting model is free of all the known loopholes of the Bell theorem, but is based on the arithmetic loophole which we will now describe in detail. 
The arithmetic perspective will lead to a product which is in between the classical multiplication from Bell-type proofs, and the tensor product from quantum mechanics. It will be quantum enough to fake quantum probabilities, and still classical enough to allow for eavesdropping over quantum communication channels. The model is meant as a proof-of-principle counterexample to Bell's theorem, and not as a full hidden-variables alternative to quantum mechanics. Neither shall we try to discuss generalizations of Bell's inequalities \cite{Zuk}, or more complicated experimental configurations \cite{ZHHH}.

Suppose that macroscopic-scale observers employ our well known Diophantine arithmetic of $\mathbb{R}$. Hidden-variable-scale theory employs some other arithmetic and calculus. However, both levels of description must agree on a probabilistic level. Accordingly, probabilities should be represented by non-negative real numbers that sum to $1$. By the term `sum' we mean here two types of addition simultaneously: the ordinary Diophantine $+$ at the  level of macroscopic observers, and some yet unspecified $\oplus$ governing the sub-quantum world. The existence of probabilities that sum to 1 in two different ways is not an entirely trivial mathematical fact.

An analogous formal structure is known from quantum mechanics, where the sum of probabilities $\sum_j p_j=1$ is compatible with the spectral sum of projectors $\sum_j \hat p_j=\hat 1$, but only the former is directly related to experiment.

For our purposes it will be enough to assume that $\mathbb{X}=\mathbb{R}$. Hidden-variable reals $\mathbb{X}$ are equipped with their own non-Diophantine sub-quantum arithmetic, and non-Newtonian sub-quantum calculus determined by a single one-to-one unknown function $f:\mathbb{X}\to \mathbb{R}$. The hidden-variable arithmetic is defined by
\be
x\oplus y &=& f^{-1}\big(f(x)+f(y)\big),\label{ar1}\\
x\ominus y &=& f^{-1}\big(f(x)-f(y)\big),\label{ar2}\\
x\odot y &=& f^{-1}\big(f(x)\cdot f(y)\big),\label{ar3}\\
x\oslash y &=& f^{-1}\big(f(x)/f(y)\big).\label{ar4}
\ee
$\oplus$ and $\odot$ are associative and commutative, and $\odot$ is distributive with respect to $\oplus$. 
$\mathbb{X}$ is ordered: $x\leq' y$ if and only if $f(x)\leq f(y)$. 
The neutral elements of addition and multiplication read, respectively, $0'=f^{-1}(0)$ and  $1'=f^{-1}(1)$.  For arbitrary real numbers $r\in\mathbb{R}$ we denote  $r'=f^{-1}(r)$. 
We will assume $0'=0$ and $1'=1$. The latter should be contrasted with quantum mechanics where $\hat 1\neq 1$.

In order to mimic the Bell construction we need the notion of an integral. Its form must be consistent with the arithmetic. We begin with the derivative, which is conceptually simpler, since once we know how to differentiate it becomes clear how to integrate.

The derivative of a function $A: \mathbb{X}\to \mathbb{X}$ is defined by (\ref{DA00})
 which, due to $0'=0$,  can be written here as 
\be
\frac{{\rm D}A(x)}{{\rm D}x}
=
\lim_{\delta\to 0}\Big(
A(x\oplus \delta)\ominus A(x)\Big)\oslash \delta,\label{DA0}
\ee

A non-Newtonian (Riemann or Lebesgue) integral is defined in a way guaranteeing the fundamental theorem of non-Newtonian calculus, linking derivatives and integrals. In particular, under certain technical assumptions paralleling those from the fundamental theorem of  Newtonian calculus, if $A$ is a function mapping a given set into itself,  $A:\mathbb{X}\to\mathbb{X}$, then  \cite{BC,GK,G79,G83,Pap1993,Pap2008,G,MC2015,ACK2016a,ACK2016b,ACK2018,Czachor2019}
\be
\int_{x_1}^{x_2} \frac{{\rm D}A(x)}{{\rm D}x}{\rm D}x = A(x_2)\ominus A(x_1)\label{ft1}
\ee
and
\be
\frac{{\rm D}}{{\rm D}x}
\int_{x_1}^x A(y){\textrm D}y 
&=&
A(x).\label{ft2}
\ee
It is easy to show that 
\be
\int_{x_1}^{x_2} A(y){\textrm D}y 
=
f^{-1}\left(\int_{f(x_1)}^{f(x_2)}f\circ A\circ f^{-1}(r){\rm d}r
\right)
\ee
where the integral over $r$ is Newtonian. 

Properties (\ref{ft1})--(\ref{ft2}) stand in contrast with other calculi one encounters in physical applications \cite{fract}, typically having great difficulties with fulfilling the fundamental theorem. The power and efficiency of the non-Newtonian approach lies in its low-level starting point --- the arithmetic. 

\begin{figure}
\includegraphics[width=5 cm]{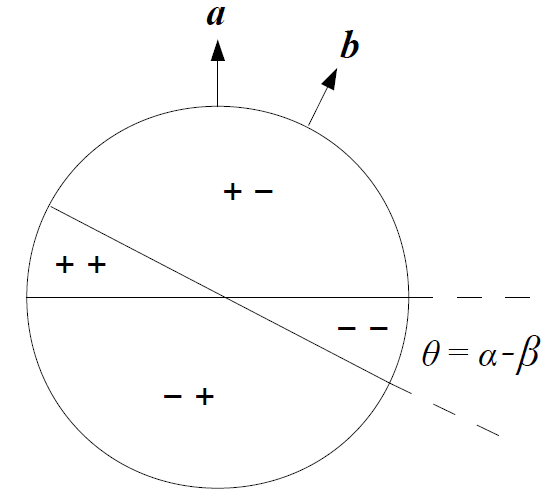}
\includegraphics[width=5 cm]{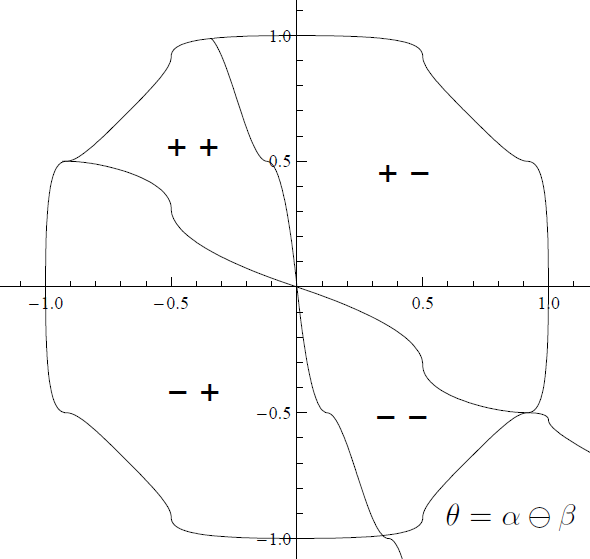}
\caption{Top: a classical model with joint probabilities $p_{++}=p_{--}=\theta/(2\pi)$, $p_{+-}=p_{-+}=(\pi-\theta)/(2\pi)$. Bottom: its non-Diophantine analogue. Despite appearances both models are rotationally invariant. In both cases the hidden variable is a point on a circle. Equivalently, hidden variables correspond to angles in, respectively, Diophantine and non-Diophantine polar coordinates.}
\label{Fig1}
\end{figure}

We just need to construct  $f$. In order to do so, consider two sets of probabilities,
\be
p'_{\pm\mp}(\theta) &=&
\frac{1}{2}\cos^2\frac{\theta}{2}\label{f11}
,\\
p'_{\pm\pm}(\theta) &=&
\frac{1}{2}\sin^2\frac{\theta}{2},\label{f12}
\ee
and
\be
p_{\pm\mp}(\theta) &=&
\frac{\pi-\theta}{2\pi}
,\label{f13}\\
p_{\pm\pm}(\theta) &=&\frac{\theta}{2\pi},\label{f14}
\ee
for $0\le\theta\le\pi$.
Obviously,
\be
1
&=&
p'_{+-}
+
p'_{++}
+
p'_{--}
+
p'_{-+}\\
&=&
p_{+-}
+
p_{++}
+
p_{--}
+
p_{-+}.
\ee
A classical model leading to joint probabilities $p_{\pm\pm}$, $p_{\pm\mp}$ is illustrated in Fig.~\ref{Fig1}. Probabilities are determined by ratios of arc lengths on a circle. The hidden variable is here given by a point on the circle or, equivalently, by its polar angle $\lambda$. Once one knows $\lambda$ the results of future measurements are known in advance. The model does not violate Bell-type inequalities. 

Our hidden-variable model will be essentially the same. We will only change arithmetic and calculus. The arc length has to be computed by means of a non-Newtonian integral, and division must be consistent with the arithmetic that defines the calculus.

Now consider the one-to-one function $f^{-1}:[0,1/2]\to [0,1/2$], defined for $0\le\theta\le\pi$
 by
\be
p'_{\pm\pm}
=
\frac{1}{2}\sin^2\frac{\theta}{2}
=
f^{-1}
\left(
\frac{\theta}{2\pi}
\right) 
=f^{-1}(p_{\pm\pm}).\label{54}
\ee
Equivalently,
\be
p'_{\pm\mp}
=
\frac{1}{2}\cos^2\frac{\theta}{2}
=
f^{-1}
\left(
\frac{\pi-\theta}{2\pi}\right)
=
f^{-1}(p_{\pm\mp}).\label{53}
\ee

Formulas (\ref{54})--(\ref{53}) might seem trivial, expressing the obvious fact that $\sin x$ is a function of $x$. What is nontrivial, however,  is that this trivial function may be nontrivially employed to construct a new  arithmetic and calculus. This is the key observation of the paper. The arithmetic will allow us to build a rotationally invariant hidden-variables model, although the notion of rotational symmetry will have to be formulated within the language of the new arithmetic.

\begin{figure}
\includegraphics[width=8 cm]{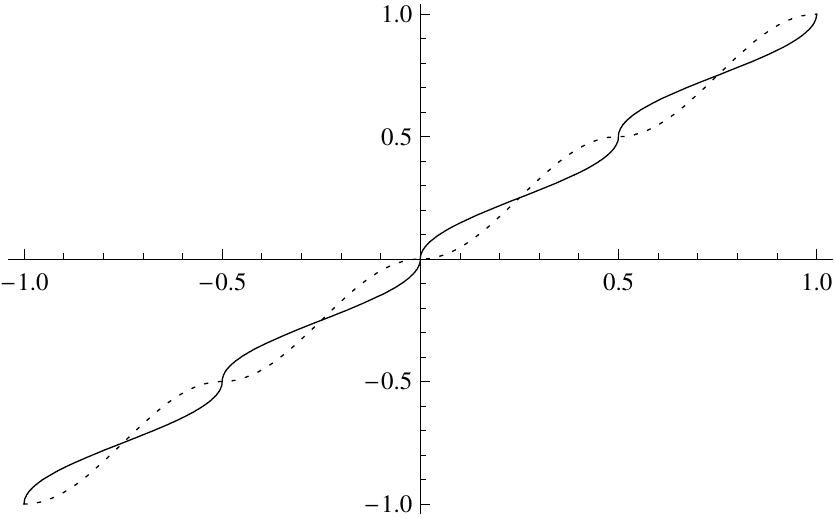}
\caption{One-to-one $f:\mathbb{R}\to \mathbb{R}$  (full) and its inverse $f^{-1}$ (dotted) defined by  (\ref{f})  and (\ref{f^-1}).}
\label{Fig2}
\end{figure}

Since (\ref{54})--(\ref{53}) are equivalent on $[0,\pi]$,  (\ref{54}) can define the restriction to $[0,1/2]$ of a one-to-one $f^{-1}:\mathbb{R}\to \mathbb{R}$. $f^{-1}(0)=0$, $f^{-1}(1/2)=\frac{1}{2}\sin^2\frac{\pi}{2}=1/2$. For example (Fig.~\ref{Fig2}),
\be
f^{-1}(x) &=&\frac{n}{2}+ \frac{1}{2}\sin^2\pi \left(x-\frac{n}{2}\right), \label{f^-1}\\
f(x) &=& \frac{n}{2}+\frac{1}{\pi}\arcsin\sqrt{2x -n}, \label{f}\\
&\pp=&
\textrm{for $\frac{n}{2}\le x\le \frac{n+1}{2}$, $n\in\mathbb{Z}$}.
\ee
The function so defined satisfies
\be
f^{-1}(n/4) = n/4 = f(n/4), \quad \textrm{for  $n\in\mathbb{Z},$}
\ee
and thus, in particular, $0'=0$, $(\pm 1)'=\pm 1$, $(1/2)'=1/2$. 
All integers are unchanged, so number theory will be unaffected. Sums, differences and products of integers are the usual ones, as opposed to their ratios.

As opposed to the harmonic arithmetic from Section~\ref{Sec2}, the non-Diophantine ordering relation $\le'$ is here identical to the Diophantine $\le$ because $f$ is strictly increasing. In consequence, $x\le' y$ if and only if $f(x)\le f(y)$, which holds if and only if $x\le y$. Modulus is thus defined in the usual way,
\be
|x|
&=&
\left\{
\begin{array}{r}
x\textrm{ if $0\leq x$}\\
\ominus x\textrm{ if $x\leq 0$}
\end{array}
\right.
,
\ee
where $\ominus x=-x$,  a consequence of $f^{-1}(-x)=-f^{-1}(x)$. 

The trigonometric identities,
\be
p'_{+-}
+
p'_{++}
&=&
p'_{-+}
+
p'_{--}
=
p'_{+-}
+
p'_{--}
=
p'_{-+}
+
p'_{++}
\nonumber\\
&=&
\frac{1}{2}\cos^2\frac{\theta}{2}+\frac{1}{2}\sin^2\frac{\theta}{2}=\frac{1}{2},
\ee
express the fact that $+$ and $-$ are equally probable.  The same is found in the hidden-variables world, although the  reasons for that are more subtle, for example,
\be
p'_{+-}
\oplus
p'_{++}
&=&
f^{-1}\big(f(p'_{+-})+f(p'_{++})\big)
\nonumber\\
&=&
f^{-1}\left(\frac{\pi-\theta}{2\pi}+\frac{\theta}{2\pi}\right)
=
f^{-1}\left(\frac{1}{2}\right)=\frac{1}{2}.\nonumber
\ee
(\ref{54})--(\ref{53}) can be rewritten as 
\be
\frac{1}{2}\cos^2\frac{\alpha-\beta}{2}
&=&
f^{-1}
\left(
\frac{f(\pi')-|f(\alpha')-f(\beta')|}{f(2')f(\pi')}\right)
\label{53'}\\
&=&
f^{-1}
\left(
\frac{f(\pi')-|f(\alpha'\ominus\beta')|}{f\big((2\pi)'\big)}\right)
\label{53''}\\
&=&
\big(\pi'\ominus|\alpha'\ominus\beta'|\big)\oslash (2\pi)',\label{p'_{+-}}\\
\frac{1}{2}\sin^2\frac{\alpha-\beta}{2}
&=&
f^{-1}
\left(
\frac{|f(\alpha')-f(\beta')|}{f(2')f(\pi')}
\right)\label{54'}\\
&=&
f^{-1}
\left(
\frac{|f(\alpha'\ominus\beta')|}{f\big((2\pi)'\big)}
\right)\label{54'}\\
&=&
|\alpha'\ominus\beta'|\oslash (2\pi)',\label{p'_{++}}
\ee
where 
\be
0\le|f(\alpha'\ominus\beta')|=|f(\alpha')-f(\beta')|=|\alpha-\beta|\le\pi,
\ee 
and
\be
\pi' &=& f^{-1}(\pi) =3+\frac{1}{2}\sin^2(\pi^2)=3.09258,\\
(2\pi)'
&=& f^{-1}(2\pi)=6+\frac{1}{2}\sin^2 (2\pi^2)=6.30175.
\ee
Probabilities (\ref{p'_{+-}}) and (\ref{p'_{++}}) are non-Diophantine ratios of arc lengths, computed by means of non-Newtonian integrals.
Indeed, the non-Newtonian integral
\be
\int_{x_1}^{x_2}{\rm D}x = \int_{x_1}^{x_2}\frac{{\rm D}x}{{\rm D}x}{\rm D}x =x_2\ominus x_1
\ee
can be used to cross-check our construction. The length of the unit circle is 
\be
\int_{0'}^{(2\pi)'} {\rm D}\lambda =(2\pi)'=f^{-1}(2\pi).
\ee
The length of the arc $\alpha'\le\lambda\le\beta'$ reads
\be
\int_{\alpha'}^{\beta'} {\rm D}\lambda ={\beta'}\ominus \alpha'=f^{-1}(\beta-\alpha).
\ee
Employing the explicit form of our hidden-variables arithmetic we obtain, for $0\leq \beta-\alpha\leq \pi$, 
\be
\int_{\alpha'}^{\beta'} {\rm D}\lambda =\frac{1}{2}\sin^2[\pi(\beta-\alpha)].
\ee
The probability of randomly selecting a point belonging to the arc is, in the hidden-variables world, the ratio of the two lengths, 
\be
&{}&
\left(\int_{\alpha'}^{\beta'} {\rm D}\lambda\right) \oslash(2\pi)'
=\left(
\frac{1}{2}\sin^2[\pi(\beta-\alpha)]
\right)\oslash(2\pi)'\nonumber\\
&\pp=&\pp=
=f^{-1}\left(
f\left(\frac{1}{2}\sin^2[\pi(\beta-\alpha)]\right)/f\big((2\pi)'\big)
\right)\nonumber\\
&\pp=&\pp=
=f^{-1}\left(
\frac{\beta-\alpha}{2\pi}
\right)
=
\frac{1}{2}\sin^2\frac{\beta-\alpha}{2}.
\ee
Notice that the ratio of lengths defines a normalized probability density
\be
\rho(\lambda)=1'\oslash (2\pi)'=\big(1/(2\pi)\big)',
\ee
with rotationally invariant normalization
\be
\int_{0'}^{(2\pi)'} \rho(\lambda){\rm D}\lambda=
\int_{\phi}^{\phi\oplus (2\pi)'} \rho(\lambda){\rm D}\lambda
=1,
\ee
for any $\phi$. 
Quantum probabilities can be thus written in terms of non-Newtonian integrals of the local-realistic form assumed in the proof of the Clauser-Horne (CH)  inequality \cite{CH} (see the next Section),
\be
p'_{++}
&=&
\frac{1}{2}\sin^2\frac{\beta-\alpha}{2}
=
\int_{\alpha'}^{\beta'} \rho(\lambda){\rm D}\lambda\label{rho1}\\
&=&
\int_{0}^{(2\pi)'} \chi_{\alpha +}^1(\lambda)\odot \chi_{\beta +}^2(\lambda)\odot
\rho(\lambda){\rm D}\lambda,\label{rho1'}\\
p'_{+-}
&=&
\frac{1}{2}\cos^2\frac{\beta-\alpha}{2}
=
\int_{\beta'}^{\alpha'\oplus\pi'} \rho(\lambda){\rm D}\lambda\label{rho2}\\
&=&
\int_{0}^{(2\pi)'} \chi_{\alpha +}^1(\lambda)\odot \chi_{\beta -}^2(\lambda)\odot
\rho(\lambda){\rm D}\lambda,\label{rho2'}\\
p'_{--}
&=&
\frac{1}{2}\sin^2\frac{\beta-\alpha}{2}
=
\int_{\alpha'\oplus\pi'}^{\beta'\oplus\pi'} \rho(\lambda){\rm D}\lambda\label{rho3}\\
&=&
\int_{0}^{(2\pi)'} \chi_{\alpha -}^1(\lambda)\odot \chi_{\beta -}^2(\lambda)\odot
\rho(\lambda){\rm D}\lambda,\label{rho3'}\\
p'_{-+}
&=&
\frac{1}{2}\cos^2\frac{\beta-\alpha}{2}
=
\int_{\beta'\oplus\pi'}^{\alpha'\oplus(2\pi)'} \rho(\lambda){\rm D}\lambda\label{rho4}\\
&=&
\int_{0}^{(2\pi)'} \chi_{\alpha -}^1(\lambda)\odot \chi_{\beta +}^2(\lambda)\odot
\rho(\lambda){\rm D}\lambda.\label{rho4'}
\ee
Here the $\chi$'s are the characteristic functions discussed below. 

As required, two normalizations hold simultaneously:
\be
p'_{++}+\dots+p'_{--}=1=p'_{++}\oplus \dots\oplus p'_{--}.\label{ie}
\ee
The right-hand form follows from the general non-Newtonian formula, for integrals of functions $F:\mathbb{X}\to \mathbb{Y}$, 
\be
\int_{a}^{b} F(x){\rm D}x
\oplus_\mathbb{Y}
\int_{b}^{c} F(x){\rm D}x
=
\int_{a}^{c} F(x){\rm D}x,\label{abc}
\ee
where $\oplus_\mathbb{Y}$ is the addition in $\mathbb{Y}$. The left-hand form guarantees that macroscopic-scale observers can test the probabilities by comparing them with experimentally measured frequencies, which necessarily sum to 1 in the arithmetic used by the observers. 

Formulas (\ref{rho1'})--(\ref{rho4'}) pinpoint similarities and differences between our hidden-variable model and those discussed in the literature so far. The difference reduces to  $\odot$ instead of `$\cdot$'. The properties of the integral are also important but it is hard to say if this is really different from what Bell had in mind. Anyway, what he assumed was that some sort of integration applies to some unspecified hidden variables.

\section{Roots of the construction: A lemma}

The origin of our construction can be traced back to the following lemma:

Consider a one-to-one $g: [0,1]\to [0,1]$. Then $g(p)+g(1-p)=1$ if and only if  
\be
g(p)=\frac{1}{2} + h\left(p-\frac{1}{2}\right),
\ee
where  one-to-one $h:[-1/2,1/2]\to[-1/2,1/2]$ is anti-symmetric.

\medskip
\noindent
{\it Proof\/}: Let $h(-x)=-h(x)$. Then
\be
g(1-p)+g(p)
&=&
\frac{1}{2} + h\left(1-p-\frac{1}{2}\right)
+
\frac{1}{2} + h\left(p-\frac{1}{2}\right)\nonumber\\
&=&
1 + h\left(\frac{1}{2}-p\right)
+
h\left(p-\frac{1}{2}\right)\nonumber\\
&=&
1 - h\left(p-\frac{1}{2}\right)
+
h\left(p-\frac{1}{2}\right)=1.\nonumber
\ee
Now, let $g(1-p)+g(p)=1$. Then,
\be
1 &=& g(1-p)+g(p)\nonumber\\
&=&
\frac{1}{2} + h\left(1-p-\frac{1}{2}\right)
+
\frac{1}{2} + h\left(p-\frac{1}{2}\right)\nonumber\\
&=&
1 + h\left(\frac{1}{2}-p\right)
+
h\left(p-\frac{1}{2}\right),\nonumber
\ee
and thus
\be
0
=
h\left(\frac{1}{2}-p\right)
+
h\left(p-\frac{1}{2}\right)
\ee
implying $h(-x)=-h(x)$. Since $g$ is one-to-one, $h$ must be one-to-one as well.
$\Box$

The lemma shows that classical and quantum formalisms are just two particular cases of an infinite family of non-Diophantine theories where a non-Diophantine probability is simultaneously a probability in the ordinary Diophantine sense of (\ref{intro}). 

The lemma is a property of binary probabilities, so it does not have a nontrivial generalization to arbitrary $n$-tuples of probabilities \cite{entropy}.
The fact that we could work with four probabilities comes from the following construction.  
Let us consider probabilities normalized by $p_1+p_2=p$. In order to  find $G$ such that 
\be
G(p_1)+G(p_2)=p
\ee
we define $q_k=p_k/p$. Then $q_1+q_2=1$ and there exists a function $g$ such that 
\be
g(q_1)+g(q_2) &=&1,\\
pg(q_1)+pg(q_2) &=&p.
\ee
Accordingly $G(p_k)=pg(p_k/p)$ satisfies 
\be
G(p_1)+G(p_2)=p, \quad p_1+p_2=p.
\ee
Two-electron singlet-state probabilities correspond to $p=1/2$, $g(q)=\sin^2\frac{\pi}{2}q$, and 
$h(x)=\frac{1}{2}\sin \pi x$.

\section{Projection postulate}

Measurement of a yes-no random variable projects  onto a subset of states corresponding to the result `yes'. In classical probability the projector is represented by a characteristic function $\chi_+(x)$, equal 1 if $x$ represents `yes', and 0 otherwise.
The orthogonal projector reads $\chi_-(x)=1-\chi_+(x)$.  In quantum probability the projection is on  a vector subspace spanned by appropriate eigenvectors. Our model is classical, so the projector is represented  by a characteristic function $\chi_\pm(\lambda)=1\ominus \chi_\mp(\lambda)$, 
\be
\chi_\pm(\lambda)\odot \chi_\pm(\lambda) &=& \chi_\pm(\lambda),\\
\chi_\pm(\lambda)\odot \chi_\mp(\lambda) &=& 0.
\ee
More explicitly, we can represent characteristic functions by the diagram
\be
\begin{array}{rcl}
\mathbb{X}                & \stackrel{\chi_\pm}{\longrightarrow}       & \mathbb{X}               \\
f{\Big\downarrow}   &                                     & {\Big\downarrow}f   \\
\mathbb{R}                & \stackrel{\tilde \chi_\pm}{\longrightarrow}   & \mathbb{R}
\end{array},\label{chi}
\ee
where,
\be
\textstyle
\tilde \chi_+(r)
&=&
\left\{
\begin{array}{l}
1 \quad \textrm{if $r'=f^{-1}(r)$ corresponds to `yes'} \\
0 \quad \textrm{otherwise}
\end{array}
\right.\nonumber,\\
\tilde \chi_-(r)
&=&1-\tilde \chi_+(r).
\ee 
Measurements reduce probability by projection and renormalization,
\be
\rho(\lambda)&\mapsto &
\rho_\pm(\lambda)\\
&=&
\chi_\pm(\lambda)\odot \rho(\lambda)
\oslash
\int \chi_\pm(x)\odot \rho(x){\rm D}x.
\ee
Joint probabilities (\ref{rho1})--(\ref{rho4'}) provide examples of the construction.

\begin{figure}
\includegraphics[width=8 cm]{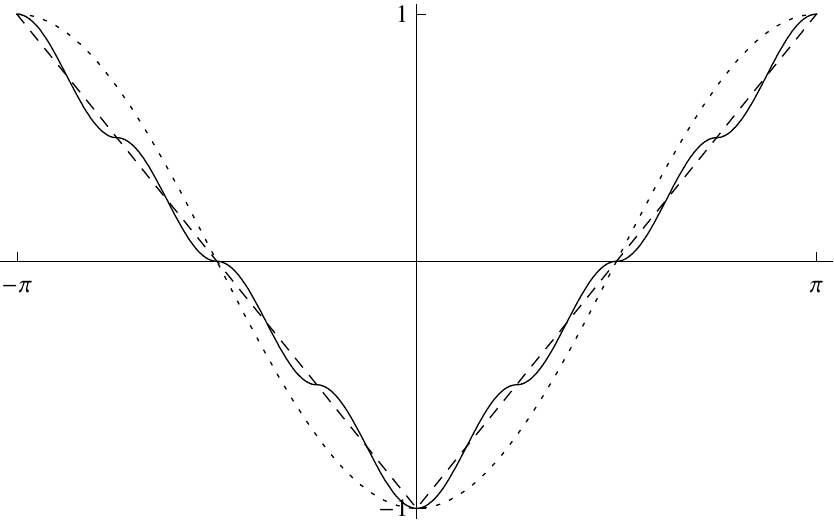}
\caption{Comparison of the three averages: $f^{-1}(-1+2|\theta|/\pi)$ (full), $-\cos \theta$ (dotted), and $-1+2|\theta|/\pi$ (dashed). Although the averages differ, the probabilities corresponding to $f^{-1}(-1+2|\theta|/\pi)$ and $-\cos \theta$ are identical. Experiments test probabilities.}
\label{Fig3}
\end{figure}

\section{Bell-type inequalities}

Bell-type inequalities are simultaneously violated and not violated, depending on the viewpoint. Let us see how it works.

\subsection{Macdonald inequality}

One of the shortest derivations of a Bell-type inequality was given by Macdonald \cite{Macdonald}  in his comment on the Pitowsky sphere model \cite{PitowskyPRL}. Let us discuss in detail the Macdonald argument from the point of view of our non-Newtonian construction. 

Let $N$ pairs of electrons, each with total spin zero, emerge in opposite directions from an interaction. Let $N(A^+:C^+)$ be the number of pairs in which the left member has spin up in the $A$ direction
and the right member has spin up in the $C$ direction. Let $N(A^+C^-:)$ be the number of pairs in which the left member has spin up in the $A$ direction and spin down in the $C$ direction. Zero total spin implies the counterfactual relation $N(A^+C^-:)=N(A^+:C^+)$.

Quantum mechanically $N(A^+C^-:)$  is an ill defined notion but we assume that it makes sense in a hidden-variable model. 
So,
\be
N(A^+:C^+) &=& N(A^+C^-:)\nonumber\\
&=& N(A^+B^-C^-:)+N(A^+B^+C^-:)\nonumber\\
&\le& N(A^+B^-:)+N(B^+C^-:)\nonumber\\
&=& N(A^+:B^+)+N(B^+:C^+).
\ee
Since for large $N$ we can write 
\be
N(A^+:C^+)\approx N p(A^+\cap C^+),\label{Np}
\ee 
we obtain a Bell-type inequality
\be
p(A^+\cap C^+) \le p(A^+\cap B^+)+p(B^+\cap C^+)
\ee
for joint probabilities. The inequality can be violated if the probabilities are quantum,
\be
p(A^+\cap B^+)=p'_{++}(\alpha-\beta)=\frac{1}{2}\sin^2\frac{\alpha-\beta}{2}.
\ee
In our model we assume that experimental data are those obtained by the macroscopic observers (since we reconstruct the measurable probabilities implied by quantum mechanics) so (\ref{Np}) must be valid as well,
\be
\frac{N(A^+:B^+)}{N}
\approx
\int\chi_{\alpha +}^1(\lambda)\odot \chi_{\beta +}^2(\lambda)\odot
\rho(\lambda){\rm D}\lambda.\label{exp data}
\ee
The left-hand side of (\ref{exp data}) is an experimental result, while its right-hand side is a hidden-variable theoretical prediction. One should {\it not\/} employ experimental data in non-Diophantine ratios of the form 
$N(A^+:B^+)\oslash N$. This is the reason why we impose the usual normalization $p'_{++}+p'_{+-}+p'_{-+}+p'_{--}=1$.

Now, let us first note that for non-Newtonian integrals
\be
\int\big(F(x)\oplus G(x)\big){\rm D}x=
\int F(x){\rm D}x
\oplus
\int G(x){\rm D}x.\label{F oplus G}
\ee
It may be not completely irrelevant to mention that the same type of generalized linearity occurs in fuzzy and Choquet integration 
\cite{fuzzy calc,fuzzy dif,Pap2020}. With this observation in mind let us repeat the Macdonald derivation. Zero-spin correlation is encoded in
\be
\frac{N(A^+:C^+)}{N}
&\approx&
\int \chi_{\alpha+}^1\odot \chi_{\gamma+}^2\odot\rho\,{\rm D}\lambda,
\nonumber\\
\frac{N(A^+C^-:)}{N}
&\approx&
\int \chi_{\alpha+}^1\odot \chi_{\gamma-}^1\odot\rho\,{\rm D}\lambda,
\nonumber
\ee
since $\chi_{\gamma-}^1=\chi_{\gamma+}^2$.
Now insert $1=\chi_{\beta-}^1\oplus \chi_{\beta+}^1$, and employ (\ref{F oplus G}),
\be
\int \chi_{\alpha+}^1\odot \chi_{\gamma-}^1\odot\rho\,{\rm D}\lambda
&=&
\int \chi_{\alpha+}^1\odot \chi_{\beta-}^1 \odot \chi_{\gamma-}^1\odot\rho\,{\rm D}\lambda
\nonumber\\
&\pp=&
\oplus
\int \chi_{\alpha+}^1\odot \chi_{\beta+}^1 \odot \chi_{\gamma-}^1\odot\rho\,{\rm D}\lambda
\nonumber\\
\label{101}
\ee
With our choice of $f$ and $f^{-1}$ the ordering relation is unchanged, so
\be
\int \chi_{\alpha+}^1\odot \chi_{\beta-}^1 \odot \chi_{\gamma-}^1\odot\rho\,{\rm D}\lambda
&\le &
\int \chi_{\alpha+}^1\odot \chi_{\beta-}^1 \odot\rho\,{\rm D}\lambda\nonumber\\
&=&
\int \chi_{\alpha+}^1\odot \chi_{\beta+}^2 \odot\rho\,{\rm D}\lambda,\nonumber\\
\int \chi_{\alpha+}^1\odot \chi_{\beta+}^1 \odot \chi_{\gamma-}^1\odot\rho\,{\rm D}\lambda
&\le&
\int \chi_{\beta+}^1 \odot \chi_{\gamma-}^1\odot\rho\,{\rm D}\lambda
\nonumber\\
&=&
\int \chi_{\beta+}^1 \odot \chi_{\gamma+}^2\odot\rho\,{\rm D}\lambda.\nonumber
\ee
Finally,
\be
{}&{}&
\int \chi_{\alpha+}^1\odot \chi_{\gamma+}^2\odot\rho\,{\rm D}\lambda
\nonumber\\
&{}&\pp=
\le 
\int \chi_{\alpha+}^1\odot \chi_{\beta+}^2 \odot\rho\,{\rm D}\lambda
\oplus
\int \chi_{\beta+}^1 \odot \chi_{\gamma+}^2\odot\rho\,{\rm D}\lambda,\nonumber\\
\label{new Mac0}
\ee
which translates into experimental frequencies as
\be
\frac{N(A^+:C^+)}{N}\le
\left(\frac{N(A^+:B^+)}{N}\right)
\oplus
\left(\frac{N(B^+:C^+)}{N}\right).\nonumber\\
\label{new Mac}
\ee
Note that the ratios in (\ref{new Mac}) are the usual Diophantine ones. 
(\ref{new Mac}) agrees with experiment, in contrast to the Macdonald inequality,
\be
\frac{N(A^+:C^+)}{N}\le
\frac{N(A^+:B^+)}{N}
+
\frac{N(B^+:C^+)}{N}\nonumber
\ee
which, of course, is not satisfied in experiment.
A simple cross-check shows that with our choice of $\oplus$
\be
p'_{++}(\alpha-\gamma)=p'_{++}(\alpha-\beta)\oplus p'_{++}(\beta-\gamma),
\ee
so there is no contradiction with (\ref{new Mac0}), (\ref{new Mac}).

Bell inequalities have been turned into a test of hidden-variable arithmetic.  It is intriguing that the non-Newtonian aspect of our derivation, namely (\ref{F oplus G}), occurs only at the counterfactual stage (\ref{101}) of the proof. One could say that actual data are processed in a Diophantine way, but the non-Diophantine aspects play a role at the counterfactual level.

\subsection{CH inequality}

The Clauser-Horne inequality \cite{CH,CH2},
\be
0\leq 3p'_{+-}(\theta)-p'_{+-}(3\theta)\leq 1,\label{CHi}
\ee
if true,  in our case would be  equivalent to 
\be
0\leq 3f^{-1}\left(\frac{\pi-\theta}{2\pi}\right)-f^{-1}\left(\frac{\pi-3\theta}{2\pi}\right)\leq 1.\label{CHii}
\ee
Notice that for $f^{-1}(x)=x$ in (\ref{CHii}) we would obtain the identity
\be
3\frac{\pi-\theta}{2\pi}-\frac{\pi-3\theta}{2\pi}=3p_{+-}(\theta)-p_{+-}(3\theta)=1,
\ee
valid for any $\theta$ and consistent with (\ref{CHii}). As expected, probabilities (\ref{f13})--(\ref{f14}) satisfy (\ref{CHi}).

Inequality (\ref{CHii}) is not valid for a large class of $f$s, but in our concrete case, setting  $\theta=\pi/4$, we find the maximal violation:
\be
&{}&
3f^{-1}\left(\frac{\pi-\pi/4}{2\pi}\right)-f^{-1}\left(\frac{\pi-3\pi/4}{2\pi}\right)
\nonumber\\
&{}&\pp=
=
3f^{-1}\left(\frac{3}{8}\right)-f^{-1}\left(\frac{1}{8}\right)=1.20711.
\ee
 Of course, (\ref{CHi}) is violated because it cannot be proved in our hidden-variable model. What {\it can\/} be proved, however, is
\be
0\leq 3\odot p'_{+-}(\theta)\ominus p'_{+-}(3\theta)\leq 1,\label{CHi'}
\ee
a fact following from (\ref{rho1'}), (\ref{rho2'}),  (\ref{rho3'}),  (\ref{rho4'}), if one follows the steps of the Clauser-Horne derivation \cite{CH}.
One can cross-check:
\be
&{}&
3\odot p'_{+-}(\theta)\ominus p'_{+-}(3\theta)
\nonumber\\
&{}&\pp=
=
f^{-1}\big[f(3) f\big(p'_{+-}(\theta)\big)-f\big(p'_{+-}(3\theta)\big)\big]
\nonumber\\
&{}&\pp=
=
f^{-1}\left(3 \frac{\pi-\theta}{2\pi}-\frac{\pi-3\theta}{2\pi}\right)
=f^{-1}(1)=1.
\ee
The model is local, deterministic, detectors are ideal, observers have free will. All the standard loopholes are absent, so Bell-type inequalities are not violated... in the non-Diophantine world of the hidden variables. The only modification is that we employ $\odot$ instead of `$\cdot$', and the integral is non-Newtonian.

\subsection{CHSH inequality}

The EPR-Bohm-Bell hidden-variables average $\langle AB\rangle'$ is computed in the hidden-variables world as follows
\be
\langle AB\rangle'
&=&
p'_{++}\oplus p'_{--}\ominus p'_{+-}\ominus p'_{-+}\\
&=&
f^{-1}\big(2|\alpha-\beta|/\pi-1\big).\label{AB'}
\ee
The average satisfies the hidden-variables  CHSH inequality \cite{CHSH},
\be
\big|
\langle A_1B_1\rangle'
\oplus
\langle A_1B_2\rangle'
\oplus
\langle A_2B_1\rangle'
\ominus
\langle A_2B_2\rangle'
\big|
\le
2.\label{hv CHSH}
\ee
The observer-arithmetic average
\be
\langle AB\rangle
&=&
p'_{++}+ p'_{--}- p'_{+-}- p'_{-+}\\
&=&
-\cos (\alpha-\beta)
\ee
nevertheless does violate the observer-arithmetic CHSH inequality. 

Fig.~\ref{Fig3} shows that (\ref{AB'}) is neither the classical average $2|\alpha-\beta|/\pi-1$ corresponding to the upper part of Fig.~1, nor the quantum one.
However, quantum experiments do not measure averages --- they measure probabilities which coincide here by construction.

\section{Hidden  rotational symmetries}

Bennett-Brassard-Mermin quantum cryptographic protocol \cite{BBM92} does not use the argument based on Bell's theorem. It directly employs one-to-one correlations in two different bases, a consequence of the rotational symmetry of singlet-state probabilities. 

So, is our model rotationally invariant? 
Yes, it is, in a subtle way. But in order to understand the subtlety,  we first have to define the action of the rotation group in $\mathbb{X}\times \mathbb{X}$, the Cartesian product of $\mathbb{X}$ with itself. We will illustrate the construction with two suggestive fractal examples.

Trigonometric functions mapping $\mathbb{X}$ into $\mathbb{X}$,
\be
\Sinn x &=& f^{-1}\big(\sin f(x)\big),\\
\Coss x &=& f^{-1}\big(\cos f(x)\big),
\ee
are periodic with the period $(2\pi)'=f^{-1}(2\pi)$ (e.g. $\Sin (x\oplus (2\pi)') =\Sinn x$).

\begin{figure}
\includegraphics[width=5 cm]{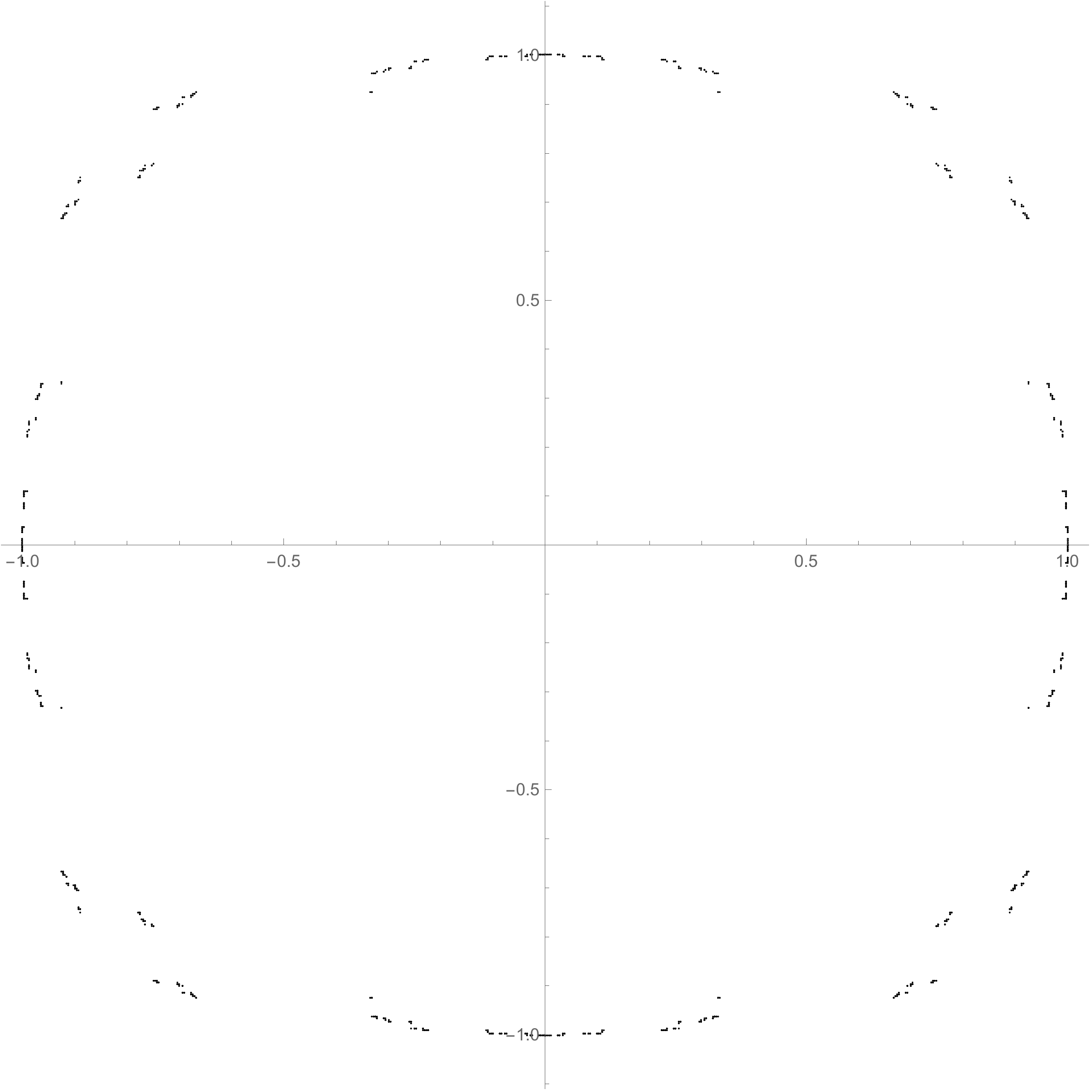}
\includegraphics[width=8 cm]{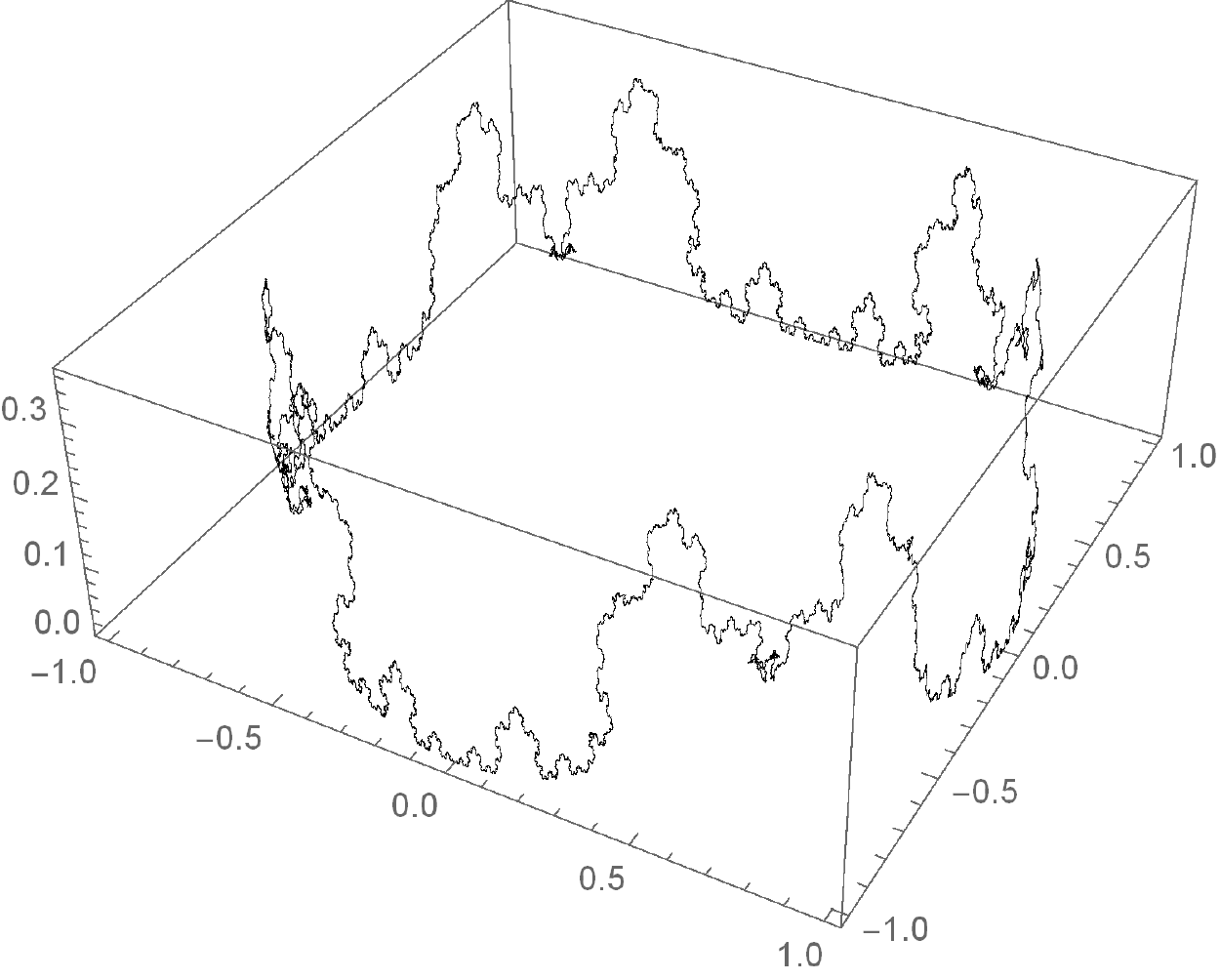}
\caption{Unit circles $\Sin^{2'}x\oplus  \Cos^{2'}x=1$ in $\mathbb{X}\times \mathbb{X}$,  where $\mathbb{X}$ is (i)  the middle-third Cantor set, and (ii) the Koch curve. Both circles are rotationally invariant in appropriate arithmetics.}
\label{Fig40}
\end{figure}

They satisfy all the standard trigonometric formulas (with respect to the arithmetic in $\mathbb{X}$), in particular:
\be
\Sin(x\oplus y)
&=&
\Sinn x\odot\Coss y\oplus\Coss x\odot\Sinn y,\label{sin xy}\\
\Cos(x\oplus y)
&=&
\Coss x\odot\Coss y\ominus\Sinn x\odot\Sinn y,\label{cos xy}\\
1 
&=&
\Sin^{2'}x\oplus  \Cos^{2'}x.
\ee
Here $\Sin^{2'}x=\Sinn x\odot\Sinn x$, etc. Rotations in the plane $\mathbb{X}\times \mathbb{X}$ are defined in the usual way, 
\be
x_1(\alpha) &=& x_1\odot \Coss\alpha \ominus  x_2\odot \Sinn\alpha,\label{rot1}\\
x_2(\alpha) &=& x_1\odot \Sinn\alpha   \oplus  x_2\odot \Coss\alpha .\label{rot2}
\ee
Formulas 
(\ref{sin xy})--(\ref{cos xy}), (\ref{rot1})--(\ref{rot2}) show that rotations form a Lie group (with group parameters subject to the non-Diophantine arithmetic). Fig.~\ref{Fig40} depicts two examples of unit circles generated by (\ref{rot1})--(\ref{rot2}): The one constructed in the Cartesian product of two Cantor sets, and the one in the Cartesian product of two Koch curves. Both circles are homogeneous spaces generated by rotations. The construction works because Cantor sets and Koch curves have the same cardinality  as the continuum $\mathbb{R}$. This is why appropriate one-to-one maps $f:
\mathbb{X}\to \mathbb{R}$ exist, and non-Diophantine arithmetics can be constructed \cite{BC,MC2015,ACK2016a,ACK2016b,ACK2018,Czachor2019}. 
The rotational symmetries from Fig.~\ref{Fig40} are `hidden' in the sense that in order to see them one must plot the curves in coordinate systems based on appropriate arithmetics. 

The property is shared by our hidden variables.

\section{Hidden rotational symmetry of the hidden-variable model}

\begin{figure}
\includegraphics[width= 7 cm]{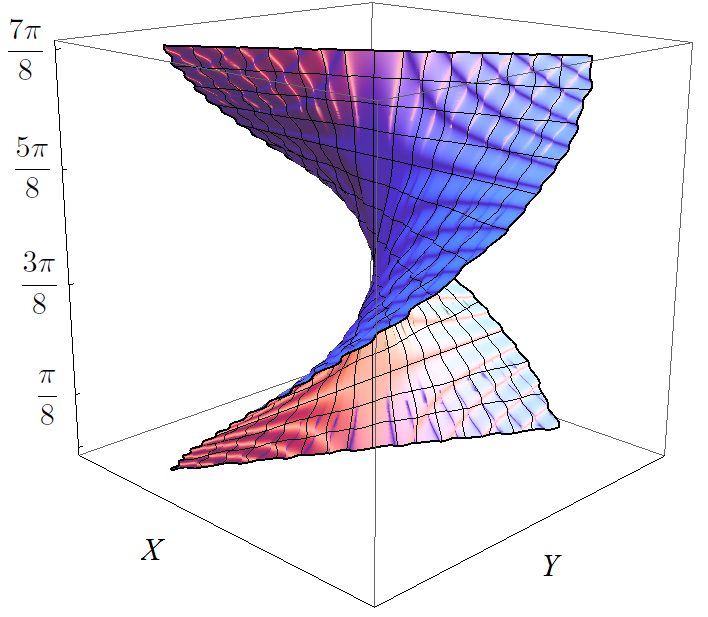}
\caption{Straight lines $t\mapsto \big(X(t),Y(t)\big)=\big(t\odot \Coss\alpha\oplus\beta,t\odot \Sinn\alpha\oplus\beta\big)$, for 
$0\leq f(\alpha)\leq 7\pi/8$, $f(\beta)=\pi/3$. Cuts through this surface for various values of $\alpha$ are shown in Fig.~\ref{Fig5}.}
\label{Fig4}
\end{figure}

\begin{figure}
\includegraphics[width=5 cm]{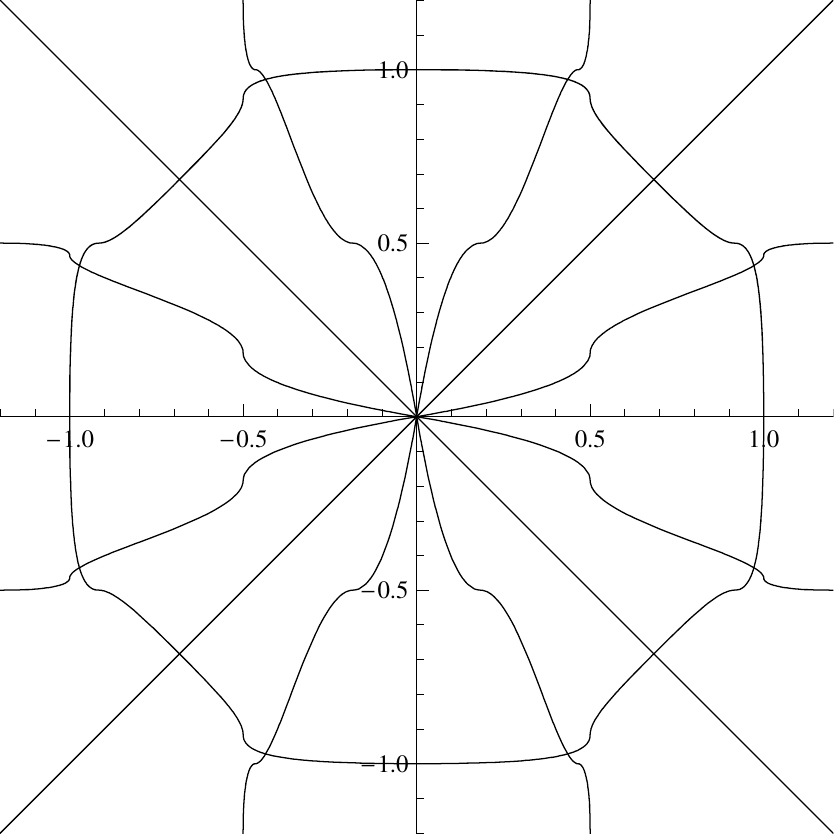}
\includegraphics[width=5 cm]{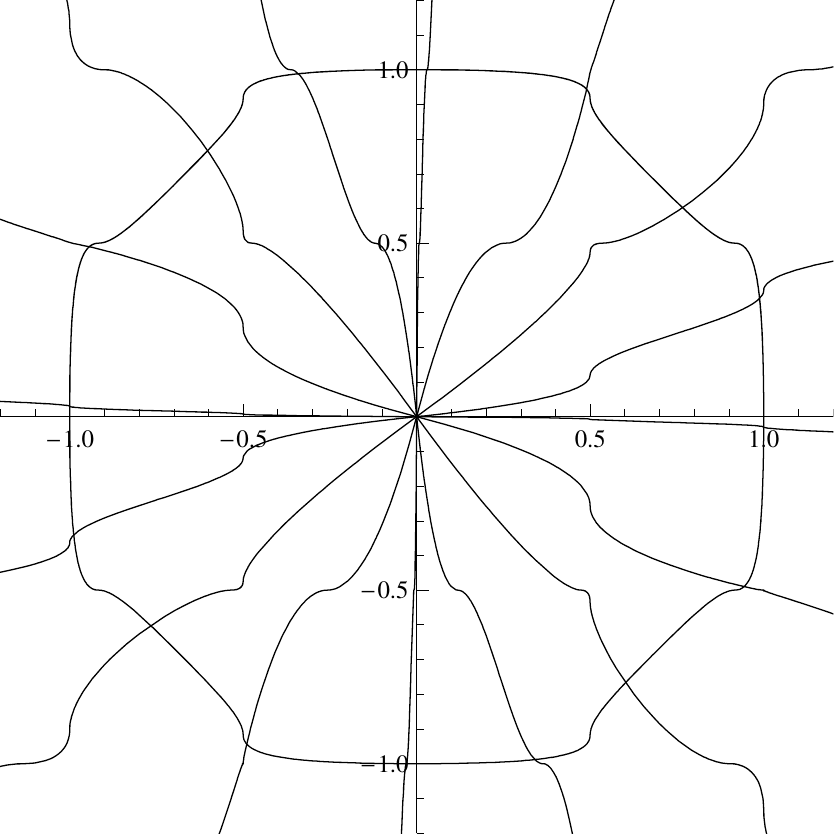}
\includegraphics[width=5 cm]{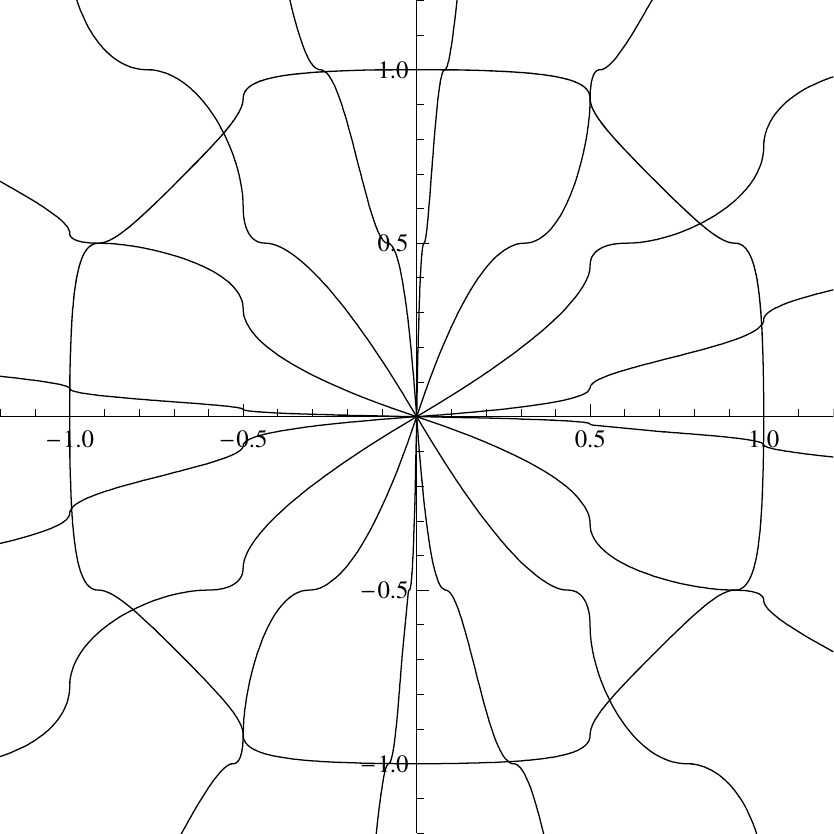}
\caption{Cuts through the surface from Fig.~\ref{Fig4} (or its rotated versions) for $f(\alpha)=n\pi/8$, $n=0,\dots,7$ and (from top to bottom) $f(\beta)=0$, $f(\beta)=\pi/10$, $f(\beta)=\pi/3$ (the latter corresponds to Fig.~\ref{Fig4}).  Non-Diophantine angular distances $\pi'\oslash 8$ between the neighboring lines are identical at all the plots. The octagon-shaped curve is the unit circle $\alpha\mapsto (\Coss\alpha,\Sinn\alpha)$, $0\le f(\alpha)\le 2\pi$.
}
\label{Fig5}
\end{figure}

Let us return to the hidden-variables arithmetic defined by (\ref{f^-1})--(\ref{f}).
A straight line through the origin is defined by (Fig.~\ref{Fig4}),
\be
t\mapsto (t\odot \Coss\theta, t\odot \Sinn\theta)
\ee
A unit circle is the curve 
\be
\phi\mapsto (\Coss\phi,\Sinn\phi), \quad 0\le\phi \le(2\pi)',\label{u-circle}
\ee 
(i.e. 
$0\le f(\phi)\le 2\pi$). In order to visualize the rotations let us draw the unit circle together with the straight lines 
$t\mapsto \big(X(t),Y(t)\big)=\big(t\odot \Coss\alpha\oplus\beta,t\odot \Sinn\alpha\oplus\beta\big)$, for 
$0\leq f(\alpha)\leq 7\pi/8$, and $f(\beta)=0$, $\pi/10$, and $\pi/3$ (Fig.~\ref{Fig5}). Non-Diophantine angular distances $\pi'\oslash 8$ between the neighboring lines are identical at all the plots. The two `deformed' plots in Fig.~\ref{Fig5} are just the rotated versions of the top one. The octagon-shaped curve is the unit circle (\ref{u-circle}). 

The circle is rotationally invariant in spite of its apparent octagon form. Needless to say, all these deformations are invisible for hidden-variable-level observers who consistently employ their own arithmetic.

\section{Further implications for cryptography}

Security of the Ekert protocol \cite{Ekert} is certified by violation of the Bell inequality. Now, which inequality: Diophantine, or non-Diophantine? Why should a  model based on Clauser-Horne type expressions (\ref{rho1'})--(\ref{rho4'}) be secure?   Knowing $\lambda$ we know in advance the results of future measurements performed by communicating parties. The model of probability we employ is internally consistent although not fully `classical'. Similarly to quantum mechanics, probabilities are constructed by means of a non-standard mathematical construction, but the end result is just a real number, an ordinary probability  that can be tested in  ordinary experiments. Probabilities sum to `one' in two ways:  Diophantine 1 and non-Diophantine $1'$ which happens to be identical to 1, but only non-Diophantine Bell-type inequalities have to be satisfied. Quantum mechanics is based on a similar structure. Probabilities sum to `one' in two ways, the ordinary 1 and the spectral $\hat 1$ for projectors, but Bell's inequality for projectors cannot be proved --- only the Tsirelson bound has to be satisfied \cite{Tsirelson}. The key difference between what we do and a Tsirelson-type reasoning is that we deal with commuting random variables only, so local incompatibility of complementary measurements is irrelevant. It might be interesting to discuss in the present context the issue of information causality \cite{IC}, but this is beyond the scope of the paper.

There are technical reasons why standard Bell-type inequalities cannot be proved. For example, (\ref{abc}) holds whatever $\oplus_{\mathbb{Y}}$ one employs, but in general
\be
\int_{a}^{b} F(x){\rm D}x
+
\int_{b}^{c} F(x){\rm D}x
\neq
\int_{a}^{c} F(x){\rm D}x,\label{abc'}
\ee
and
\be
\int_{a}^{b} [F(x)+G(x)]{\rm D}x
\neq
\int_{a}^{b} F(x){\rm D}x
+
\int_{a}^{b} G(x){\rm D}x,\label{abc''}
\ee
so all the proofs \`a la Bell one finds in the literature will not work, as we have explicitly seen in the Macdonald inequality example. Non-Newtonian integrals are linear maps but with respect to appropriate non-Diophantine arithmetic. With respect to the Diophantine arithmetic they are nonlinear. This type of duality is well known in physics (nonlinear waves interfere, $n$-soliton solutions are formed by Darboux-B\"acklund transformations from 1-soliton solutions, Lax-pair represents a nonlinear system by a linear one, etc.).  
If one consistently works according to the non-Diophantine/non-Newtonian rules, Bell-type inequalities can be proved, but their correct form is exemplified by (\ref{CHi'}) and (\ref{hv CHSH}), and not by (\ref{CHi}) or the like.

\section{Summary}

All the papers on Bell's theorem begin (either explicitly or implicitly) with probabilities of the form
\be
\int p^1_{\alpha}(\lambda)p^2_{\beta}(\lambda)
\rho(\lambda){\rm d}\lambda .\label{S1typ}
\ee
Here $\lambda$s are some unspecified  hidden variables, $\rho(\lambda){\rm d}\lambda$ is an arbitrary probability measure, $\int$ is the integral associated with the measure. Probabilities can be added so that the overall probability is normalized to 1. 
The product $p^1_{\alpha}(\lambda)p^2_{\beta}(\lambda)$ reflects the fact that measurements depend locally on some parameters $\alpha$ and $\beta$ controlled in separate measuring procedures. 

Apparently the construction is completely general. However, we point out that: (i)  $p^1_{\alpha}(\lambda)p^2_{\beta}(\lambda)$ involves some concept of a product, and (ii) $\int$ is based on some notion of a sum. Linearity of the integral is implicitly linked with the forms of multiplication and addition employed in its construction. The sum implicitly present in the integral should be nevertheless consistent in some way with the sum employed in experiment. If our ambition is to eliminate {\it all\/} local-realistic theories, the arithmetic aspects should not be overlooked. 

Our construction satisfies all these desiderata, we just make products and sums explicit. In particular, we construct singlet-state probabilities (\ref{rho1})--(\ref{rho4'}) as follows,
\be
p'_{jk}(\beta-\alpha)
=
\int_{0}^{(2\pi)'} \chi_{\alpha j}^1(\lambda)\odot \chi_{\beta k}^2(\lambda)\odot
\rho(\lambda){\rm D}\lambda,\label{S1}
\ee
where the abstract ${\rm d}\lambda $ is replaced by a concrete non-Newtonian ${\rm D}\lambda$, and the parameters are related by $\alpha'=f^{-1}(\alpha)$, $\beta'=f^{-1}(\beta)$, 
$\alpha,\beta\in[-\pi,\pi]$, $\alpha',\beta'\in[-\pi',\pi']$. 

Products  and integrals at both sides of (\ref{rho1})--(\ref{rho4'}) are defined by means of arithmetic operations from two  different arithmetics, both acting in $\mathbb{R}$: The observer-level Diophantine arithmetic  $\{\mathbb{R},+,-,\cdot,/,\le\}$, and the hidden-variable-level non-Diophantine arithmetic $\{\mathbb{R},\oplus,\ominus,\odot,\oslash,\le\}$.

Setting $f(x)=x$ one reconstructs  Bell-type inequalities discussed in the literature, but for the price of a model that does not agree with experiment. Standard local hidden-variable theories have been turned into unphysical Newtonian special cases of a more general, local and deterministic  non-Newtonian theory.

Formulas such as (\ref{rho1})--(\ref{rho4'})  make sense because both arithmetics act in the same set $\mathbb{R}$. Since unit elements in both arithmetics are the same, $1'=1$, the probabilities are normalized in two coinciding ways. The observer-level normalization,
\be
p'_{++}+p'_{+-}+p'_{-+}+p'_{--}=1,
\ee
and the hidden-variables normalization,
\be
p'_{++}\oplus p'_{+-}\oplus p'_{-+}\oplus p'_{--}=1'=1.
\ee
This should be contrasted with quantum mechanics where two resolutions of unity coexist as well, but with different unit elements: the real number $1$, and the unit operator $\hat 1$. 

Probabilities (\ref{rho1})--(\ref{rho4'}) have a geometric representation: they represent non-Diophantine ratios of arc lengths on the unit circle $\Sin^{2'}x\oplus  \Cos^{2'}x=1$. The set of hidden variables is just the unit circle (which can be identified, if one wishes, in the usual way with its covering space $\mathbb{R}$ equipped with non-Diophantine arithmetic). Both the circle itself, and the probabilities are rotationally invariant. The latter explicitly follows from 
\be
\alpha'\ominus\beta'=
(\alpha'\oplus\phi)\ominus(\beta'\oplus\phi)
\ee
for any $\phi\in \mathbb{R}$.

\section{Concluding remarks}

The arithmetic perspective presented in this manuscript creates some `wiggle room' between the classical modeling and quantum mechanics.  Exactly how much room is available  for those who try to hack the quantum-encrypted systems remains to be studied, but several remarks are in place. 

First of all, let us stress again that the goal  was not to propose a complete hidden-variable alternative to quantum mechanics. Our objectives are minimalistic. We just show that the argument of Bell can be circumvented by making explicit a point that was overlooked in the original construction. The loophole is of fundamental origin so cannot be fixed by technological developments.

Secondly, one should study along similar lines all the known cryptographic protocols. The Bennett-Brassard protocol \cite{BB} is known to be insecure if hidden variables exist \cite{ACP}. If Bohmian hidden variables are realized in Nature then Ekert \cite{Ekert} and  Bennett-Brassard-Mermin protocols \cite{BBM92} require modifications \cite{ACP}. 

Concerning the latter, what about the arithmetic loophole in entangled-state protocols that are not directly based on Bell's theorem? Can we fake quantum correlations by an appropriate choice of arithmetic or calculus? These are open questions, but one should not be overoptimistic. Non-Diophantine and non-Newtonian methods are very flexible. They can easily mimic typically `quantum' features such as incompatible random variables or maximal sets of simultaneously measurable physical quantities \cite{Czachor2020}. 

Hackers, as opposed to Nature, are clever and malicious. Should we worry? Perhaps yes. As A.~Ekert has once expressed, `among those who make a living from the science of secrecy, worry and paranoia are just signs of professionalism'  \cite{EkertUW,EkertRenner}.

\acknowledgments

I'm indebted to Krzysztof G\'orny (Mayo Clinic), Ryszard Horodecki (UG), Maciej Kuna (PG), Marian Kupczy{\'n}ski (UQu\'ebec), {\L}ukasz Rudnicki (UG), and Marek \.Zukowski (UG)  for comments. Calculations were carried out at the Academic Computer Center in Gda{\'n}sk. The work was supported by the CI TASK grant `Non-Newtonian calculus with interdisciplinary applications'.

\end{document}